\documentclass{aa}  

\usepackage{graphicx}
\usepackage{txfonts}
\usepackage{lipsum}
\usepackage{subcaption}       
                                
\usepackage{lscape}             
\usepackage{placeins}         

\usepackage{natbib}
\usepackage[breaklinks=true]{hyperref} 
\bibpunct{(}{)}{;}{a}{}{,}  
\usepackage{xcolor}

\DeclareUnicodeCharacter{2212}{-}
\usepackage{newunicodechar}
\newunicodechar{≤}{\ensuremath{\leq}}

\begin{document}
\let\linenumbers\relax

   \title{Numerical investigation of particle acceleration at interplanetary shocks: diffusive and superdiffusive scenarios}

   \author{G. Prete\inst{1}\fnmsep\thanks{Corresponding author: giuseppe.prete@unical.it}
        \and G. Zimbardo\inst{1,2}\fnmsep
        \and S. Perri\inst{1,2}
        }

   \institute{Department of Physics, University of Calabria, Ponte P. Bucci, Cubo 31C, Rende, 87036, Cosenza, Italy.
   \and National Institute for Astrophysics, Scientific Directorate, Viale del Parco Mellini 84, I-00136 Roma, Italy.}

   \date{xxx}

\abstract  
{Energetic particles are ubiquitous in space and astrophysical plasmas, and interplanetary shocks are widely regarded as one of the main particle accelerators in the heliosphere. Indeed, in-situ measurements typically show that energetic particle fluxes peak at the shock, indicating a local acceleration process. Furthermore, the time profile of energetic particle fluxes is highly influenced by particle transport properties upstream and downstream of the shock.}
{By advancing previous numerical test-particle models that simulate the transport of monoenergetic particles around an infinite planar shock, in this work we add the acceleration of such particles via energy gains at each shock crossing, in a first-order Fermi-type mechanism.} 
{Moreover, the acceleration of a 70 keV particle population, namely the seed population, is reproduced by integrating a Langevin-type equation upstream and downstream of an infinite planar shock. 
Particles can diffuse in the simulation box via random ``kicks”, which belong either to a Gaussian distribution (normal diffusion) or to a Lévy distribution (superdiffusion).
We perform several  simulations by varying the parameters of the model.} 
{The particle energy spectra in both diffusive and superdiffusive simulations are in remarkable agreement with the theoretical predictions.The output energetic particle densities have been compared with those observed by the ACE spacecraft during an interplanetary shock crossing on December 14, 2006.}
{We show not only that particle fluxes in different energy bins reproduce very well the observed ones upstream and downstream when superdiffusion is at work, but also that anomalous, superdiffusive transport speeds up the acceleration process and leads to values of particle energies consistent with observations.}

   \keywords{Interplanetary Shocks --
                Diffusive Shock Acceleration --
               Anomalous diffusion --
               Energetic Particles
               }

   \maketitle

\section{Introduction}
One of the most efficient ways to accelerate particles in different astrophysical environments is shock acceleration \citep{Blandford78,bell1978acceleration,Kirk99,Zank07,Lee12,Helder12}. 
It is widely accepted that magnetic field fluctuations, both pre-existing in the medium and self-generated by particle streaming instabilities, as well as magnetic islands and turbulent structures, can scatter energetic particles back and forth across the shock, leading to efficient acceleration \citep{Bell04,Caprioli14a,Caprioli14b,Leroux15,Zank15}.
Numerical simulations of shock waves propagating in a turbulent medium have shown how upstream fluctuations and structures, locally generated by the turbulent flow, can lead to a strong modification of the front \citep{Giacalone08,Nakanotani21,Trotta21,Prete25,Trotta25,Zhao25}; this influences particle acceleration together with self-generated turbulence close to the shock \citep{Ha25}. On the other hand, the presence of pre-existing turbulence in astrophysical shocks, such as interplanetary (IP) shocks, the terrestrial bow shock, and shocks formed in supernova remnants, has widely been witnessed by both in situ and remote spacecraft observations \citep{Reynoso13,Pitna21,Sorriso21,Zhao21,Rakhmanova22,Trotta22,Ferrazzoli23,Perri23,Mercuri25}. Further, the non-stationarity of the shocks and the evolution of the environment conditions can also lead to shock distortions in a quasi-perpendicular configuration in terms of shock reformation and ripples, as observed at the Earth bow shock \citep{Lobzin07,Johlander16}, influencing electron and ion propagation and transmission across the shock.

In-situ spacecraft measurements have revealed that energetic particle intensities around the interplanetary shock fronts can have different spatial and temporal shapes, both related to the local shock conditions, to timing effects, background conditions, and acceleration mechanisms at work. In particular, \citet{Kartavykh25}, by analyzing more than $50$ IP shock crossings and investigating the electron and the ion intensities, detected different responses in the proton fluxes in several energy channels and deduced that only a small fraction of IP shocks are able to accelerate electrons, mainly via shock drift acceleration. In addition, the shock parameter that mostly influences the shapes of the proton fluxes is the shock speed: for velocities greater than $800$ km/s the response tends to recover the standard profile predicted by the diffusive shock acceleration (DSA), namely an upstream exponential rise towards the shock front and a constant flux downstream \citep{Drury83, Giacalone12}; conversely, for lower values of the speed, weak responses or more peaked proton profiles have been observed. 
In this connection, \citet{Perri07,Perri08,Perri09} have shown that energetic electrons and ions accelerated at IP shocks at 1 au and in the outer heliosphere can be characterized by a flux that rises as a power law from far upstream towards the shock front and that decreases slightly downstream. Such observations have been interpreted and modeled by invoking anomalous, superdiffusive transport of particles upstream of the shock \citep{Perri12a}. Superdiffusion implies a particle propagation faster than the classical Brownian-like motion, i.e., $\langle \Delta \mathbf{r}^2\rangle\propto t^{\alpha}$, with $\alpha>1$, and  \citet{Perri07,Perri08} have shown that the exponent $\alpha$ is indeed related to the exponent of the far upstream power-law particle flux, so that measuring particle fluxes can lead to the determination of the particle transport properties. 

Recently, \citet{Effenberger25} presented a review on anomalous transport in space and astrophysical plasmas, where several approaches to anomalous diffusion are discussed and particle acceleration is also considered. 

One of the key questions to clarify is related to the shock and the environmental conditions that can favor shock acceleration with superdiffusive particle transport. This is indeed a pivotal point, since \citet{Perri15} have shown that in such a case the acceleration times of the process can be shorter than in the case of standard DSA. It is important to keep in mind that the formalism used to derive the power-law particle fluxes was based on L\'evy walks \citep{Klafter87,Zumofen93,Zaburdaev15}, where the free path length and the travel time are coupled (they are not independent as in the L\'evy flight formalism--see a discussion in Section \ref{sec:energyspectra}), so that very long displacements require long times to be performed by the particle. L\'evy walks represents a more realistic description of the particle dynamics than L\'evy flights. Further, the inclusion of a superdiffusive transport in a first order Fermi-like acceleration process \citep{fermi1949origin} modifies the particle energy spectra. Indeed, \citet{Perri12a,Zimbardo13} have shown  that the slope of the energy spectra depends not only on the gas compression ratio of the shock, as in DSA, but also on the exponent $\alpha$ of the mean square displacement; this leads to harder spectra for relativistic and non-relativistic particles. Notice that particle energy spectra harder than predicted by DSA are observed not only at IP shocks \citep{vanNes84,Decker08}, but also at supernova remnants \citep{Whiteoak96, Helder12} and at shocks formed by the merging of galaxy clusters \citep{vanWeeren17,Zimbardo17,Zimbardo18}.

The effects of particle superdiffusion at the shock have been investigated by means of a test particle numerical code based on  the integration of a Langevin-type equation for an ensemble of monoenergetic test particles, where they undergo a random walk upstream and downstream of an infinite planar shock in terms of L\'evy walks \citep{Prete2019}. The scattering times (modeled as random kicks) have a typical value and are related to the particle mean free path in the case of normal transport, while they are distributed as a power-law in the case of superdiffusion in order to implement the scale-free nature of the free path lengths (notice that for superdiffusion the mean free path diverges \citep{Klafter87,Zimbardo13,Zaburdaev15}). The results of those numerical experiments compare very well with energetic ion fluxes for an appropriate  range of the model parameters, which gives strength to the anomalous transport of energetic particles in the interplanetary medium \citep{Prete2019,prete2021energetic}. Recently, \citet{Effenberger24,Aerdker25} approached the problem of particle acceleration at IP shocks via a L\'evy flight model that  reproduces both the typical power-law particle profile upstream of the shock front and particle energy spectra harder than those predicted by DSA and by a L\'evy walk model \citep{Perri12a}. Further, \citet{Aerdker25} also found an acceleration faster than in DSA, in agreement with \citet{Perri15}. 
Building on our previous studies, we present here test-particle numerical simulations that also include momentum gains when particles cross the shock front. The aim is to validate the energy spectral index and the acceleration time of superdiffusive shock acceleration in a parameter range appropriate to spacecraft IP shock crossings, and to compare with simulations of shock acceleration in the case of normal diffusion. The numerical results will be finally directly compared with an IP shock observed by the ACE spacecraft at 1 au.

\section{Numerical model and set-up}\label{sec:numerical}

Here we describe a new implementation of the 1D numerical model presented in \citet{Prete2019,prete2021energetic}. We set up two different numerical simulations: one that includes diffusive transport of energetic particles, and another one where particles can move according to a L\'evy random walk (i.e., superdiffusion). The simulation box is devised according to IP shock observations and extends from $L=-10^{10}$ m to $L=10^{10}$ m, while particles are injected at $x=0$, which coincides with the position of the shock (in the shock rest frame). The plasma  velocity of the upstream side is indicated by $V_1$, while the velocity of the downstream side is $V_2$. 
The simulation box size is set to be finite, since IP shocks propagating from the Sun to the near-Earth environment represent finite systems; still, the simulation box is large enough to minimize border effects that include the loss of high energy particles, especially within the L\'evy walk scenario, where particles are allowed to perform long free paths.

In order to mimic the propagation of particles accelerated at shock waves, which is a combination of diffusive motion, due to the scattering of particles from magnetic irregularities, and ordered plasma bulk motion, we integrate a Langevin-type equation \citep[see][]{Prete2019,prete2021energetic}
\begin{equation}
    dx_i = (V_{bulk} + v_{random})dt_i,
    \label{eq:langevin}
\end{equation}
where $V_{bulk}$ indicates the plasma speed in both the upstream and downstream regions, namely $ V_{bulk}=V_1$ if $x \leq 0$ and $V_{bulk}=V_2$ if $x > 0$. The scattering of particles by magnetic irregularities is thoroughly modeled by random kicks via the particle velocity  $v_{random}$ 
\begin{equation}
    v_{random}=(2 \xi-1)v  \quad {\rm where} \: \xi \in [0,1] \ , 
    \label{eq:vrand}
\end{equation}
with $v$ the particle speed and $\xi$ representing an uncorrelated random number that changes after a scattering time $\tau_s$. Notice that this process can be assimilated to particle motion along the magnetic field in the presence of a parallel shock, i.e., with the average magnetic field parallel to the shock normal, so that the term $2\xi -1$ in Eq.~(\ref{eq:vrand}) corresponds to the particle pitch angle cosine.\footnote{To avoid confusion with the power-law exponent $\mu$ introduced below, here we do use the standard notation of $\mu$ for the pitch-angle cosine.} Given that $\xi$ is evenly distributed between 0 and 1, Eq. (\ref{eq:vrand}) implies an isotropic particle distribution \citep[e.g.,][]{Lasuik17,Zim20}. \\
The scattering time $\tau_s$ is chosen in different ways depending on whether we are studying normal diffusion or superdiffusion. When normal transport is modeled in the simulation, the particle random displacement $\Delta x$ is related to $\tau_s$ via the particle speed as
\begin{equation}
\Delta x = v_{random} \, \tau_s .
\label{eq:tau_scatt}
\end{equation}
Since $\tau_s$ and $v_{random}$ have both finite variances, this implies that $\Delta x$ is finite, too, and does not depend on the displacement at previous times. 

When superdiffusion is implemented in the model, the particle motion is described by a L\'evy random walk. This involves a probability $\Psi$ of free paths $\ell$ with a power-law shape \citep{Zimbardo13,Trotta15, Prete2019}
\begin{equation}
\Psi (\ell, \tau_s) =
\left\lbrace
\begin{array}{ll}
\frac{1}{2}C\delta(|\ell| - v\tau_s), \;\;\;\;\;\;\;  |\ell| < \ell_0 \\
\\
\frac{1}{2}C |{\ell}/{\ell_0}|^{-\mu}\delta(|\ell| - v\tau_s), \;\;\;\;\;\;\;  |\ell| > \ell_0 \, .
\end{array}
\right.
\label{eq:levyWalk}
\end{equation}
Here $\ell_0$ indicates a scale parameter, namely a characteristic free path length, such that for $\ell>\ell_0$ the probability distribution of the path lengths is a power-law decay with exponent $\mu$. Notice the presence of the Dirac delta in the expression of $\Psi(\ell, \tau_s)$, which ensures the coupling between free path lengths and scattering times $\tau_s$. The exponent $\mu$ quantifies the weight of the power-law tails of $\Psi(\ell, \tau_s)$, with smaller $\mu$ implying  heavier tails.  
For superdiffusion the mean square displacement of particles assumes the following form, 
\begin{equation}
\langle \Delta x^2 \rangle \propto t^\alpha
\label{eq:meanSquareDispl}
\end{equation}
for $t \to \infty$, with the exponent $\alpha = 4 - \mu$ for $2 < \mu < 3$; on the other hand, normal diffusion is restored for $ \mu \geq 3$. Given that the normalization condition is fulfilled
\begin{equation}
    \int_{0}^{+\infty} d\tau_s \int_{-\infty}^{+\infty} d\ell \: \Psi(\ell,\tau_s)=1,
\label{eq:integraleNorm}
\end{equation}
we are able to define the scattering times  for a L\'evy walk process in terms of a random number $\xi$ for $\xi>\xi_0$ as 
\begin{equation}
    \tau_s = \tau_0 \left[\frac{1}{ \mu \left(1 - \xi \right)}\right]^\frac{1}{\mu - 1} ,
    \label{eq:scattSSA}
\end{equation}
while $\tau_s=\xi/C$ for $\xi < \xi_0$ (for more details see \citet{Prete2019,prete2021energetic}  and the Appendix below). Here, $\xi_0=C\,\tau_0$ separates the domain where $\psi(\tau)$ is constant from that where it is a power law.  Because of the $\delta$-coupling in Eq. \ref{eq:levyWalk}, $\tau_0$ is related to $\ell_0$ by $\tau_0 = \ell_0 / v$, where $v$ is the particle speed.
When $\xi \to 1 $, very long $\tau_s$ are allowed, and due to the coupling between space and time Eq.  (\ref{eq:scattSSA}) leads to the emergence of very long displacements.
In summary, the main difference between the two versions of the code is related to the form of the scattering times, which in the DSA case has a single typical value, while in the superdiffusive case $\tau_s$ have a power-law distribution depending on the parameters $\mu$ and $\tau_0$. 

Beyond the implementation of the particle transport as in  \citep{Prete2019}, here the novelty stays in the fact that particles can gain momentum as they cross the shock front in a Fermi-like process. 
The particle momenta in the upstream/downstream frames are related to the ones in the shock frame via a Galileian transformation. The modulus of the particle momentum is given by $p_i=p_{\rm sh} (1-\cos(\theta) V_i/V_{\rm sh})$, being the index $i=1,2$ if particles are either in the upstream or in the downstream frame, respectively. Thus, the variation in momentum when a particle goes from upstream to downstream, is given by 
\begin{equation}
p_{1\rightarrow 2} = p_1 [1+\cos(\theta)  (V_1-V_2)/v_1],
\label{eq:momentum12}
\end{equation}
being $v_1$ the particle speed in the upstream frame, and the pitch angle cosine $\in [0,1]$. On the other hand, the variation in momentum when a particle goes from downstream to upstream, is given by 
\begin{equation}
p_{2\rightarrow 1} = p_2 [1+\cos(\theta)  (V_2-V_1)/v_2],
\label{eq:momentum21}
\end{equation}
being $v_2$ the particle speed in the downstream frame, and $\cos(\theta) \in [-1,0]$.

Following \citet{Drury83}, the momentum variation for a particle passing from upstream to downstream and viceversa is the same. Thus, the average momentum gain (averaged over all possible pitch angles $\theta$, under isotropic conditions) while crossing the front is given by
\begin{equation}
\langle \Delta p\rangle=\frac{2}{3} p (V_1-V_2)/v;
\label{eq:momentum}
\end{equation}
with $p$ and $v$ the momentum and speed of particles in the upstream fluid frame, respectively. The particles’ velocities are first evaluated in the upstream and downstream reference frames and then transformed into the shock reference frame via a Galilean transformation.

We assume that particles have already been pre-energized by processes like shock drift acceleration and/or multiple shock reflections \citep[e.g.,][]{Trattner23}, and we inject particles with an initial energy of $E_0$= 70 keV, so that they can go into a first order Fermi process. We start from this energy value because it is one of the lowest-energy channels of the energetic particle instruments onboard satellites such as ACE and WIND, and we investigate which higher-energy channels are populated. Since particles are accelerated each time they cross the shock, and since we aim to compare the results from the model with the data measured by satellites, we binned particles into five energy channels, namely 100-200 keV, 200-300 keV, 300-400 keV, 400-500 keV, and $>$ 500 keV. We fixed the simulation end time to 10$^5$ s, which is larger than one day and is of the order of the typical lifetime of fast interplanetary shocks.

\section{Results}
\subsection{Parametric study of diffusive and superdiffusive shock acceleration} 
We perform two sets of numerical runs in which test particles can cross the shock both diffusively and superdiffusively. Table \ref{tab:Table1} shows the parameters used in the simulations. We recall that the compression ratio $r$ is defined as the ratio between the fluid speed upstream and the fluid speed downstream in the shock rest frame; values for the injection particle energy $E_0$, for $V_1$ and $V_2$, and for $r$ are taken from IP shock observations \citep[e.g.,][]{Giacalone12}. In the case of normal diffusion, the acceleration time depends on the diffusion coefficients upstream and downstream of the shock \citep[e.g.,][]{Drury83}, and the scattering time is directly proportional to the parallel diffusion coefficient $D_\parallel$
 via the particle speed as
\begin{equation}
    \tau_s \simeq \frac{3 D_{\parallel}}{v^2} .
    \label{eq:scattering_times1}
\end{equation}
Since $D_{\parallel}$ can be obtained from the component of the diffusion tensor along the radial direction (along the normal to the shock), $D_{xx}$, namely $D_{\parallel} = D_{xx}/\cos^2 \theta_{Bn}$, one can infer the scattering time directly from the analysis of the upstream  exponential rise of energetic particle fluxes, $J\propto \exp{[V_1 x/D_{xx}]}$ for $x<0$. As an example, we can consider that analyzing a set of three strong ACE shocks, \citet{Giacalone12} found radial diffusion coefficients ranging from $0.6\times 10^{18}$ cm$^2$ s$^{-1}$ to $3.1\times 10^{18}$ cm$^2$ s$^{-1}$ for energetic protons in the energy channels from 47 keV to 310 keV. Considering the corresponding shock normal angles $\theta_{Bn}$  and the average velocities in each energy channel, scattering times ranging from 35 s to 150 s can be inferred. Another example is given by \citet{Ding24}, who, analyzing the September 6, 2022 shock crossing measured by Solar Orbiter, obtained radial diffusion coefficients very close to the shock in the range from $5\times 10^{18}$ cm$^2$ s$^{-1}$ to $1.8\times 10^{19}$ cm$^2$ s$^{-1}$ for energetic protons with energies from 300 keV to 1 MeV. These coefficients lead to scattering times from about 60 s to 75 s. Recently, for energetic protons accelerated at shocks observed by Solar Orbiter,  \citet{Kartavykh25} found mean free paths  of the order of 0.013--0.015 au for particles of 130 keV or 1 MeV. The corresponding scattering times are of the order of 140 s. 
For the following runs, we set $\tau_s$ to 25 s, 50 s and 100 s, for the diffusive case. 

In case of superdiffusion, \citet{prete2021energetic} found for some ACE shock crossings,  that the parameter $\tau_0$, related to the scattering times via eq. (\ref{eq:scattSSA}), range from about 1 s to 5 s (the methodology is outlined in Section 3.3 below). Here, we fix $\tau_0=2$ s, 5 s and 10 s. 
It should be born in mind that superdiffusion,  owing to its scale free nature, encompasses both long scattering times (much longer than $\tau_0$) and short scattering times.

\begin{table*}[h!]
\caption{DSA vs SSA simulations parameters.}
\label{tab:Table1}
\centering
\resizebox{\textwidth}{!}{
\begin{tabular}{c c c c c c c c c c}
\hline\hline
 & n° particles & E$_0$(keV) & V$_1$(km/s) & V$_2$(km/s) & r & t$_f$(s) & $\tau_s$(s) & $\tau_0$(s) & $\mu$ \\
\hline
DSA & 10$^7$ & 70 & 400/600 & 200 & 2/3 & 10$^5$ & 100/50/25 & - & - \\
\hline
SSA & 10$^7$ & 70 & 400/600 & 200 & 2/3 & 10$^5$ & - & 10/5/2 & 2.5\\
\hline
\end{tabular}
}
\tablefoot{
Parameters adopted in the simulations with normal diffusion (DSA) and  with superdiffusion (SSA). Values separated by a slash refer to different simulation setups (e.g., $V_1 = 400$ km/s and $V_1 = 600$ km/s).

}
\end{table*}

\begin{figure}[ht!]
\centering
\includegraphics[width=9cm]{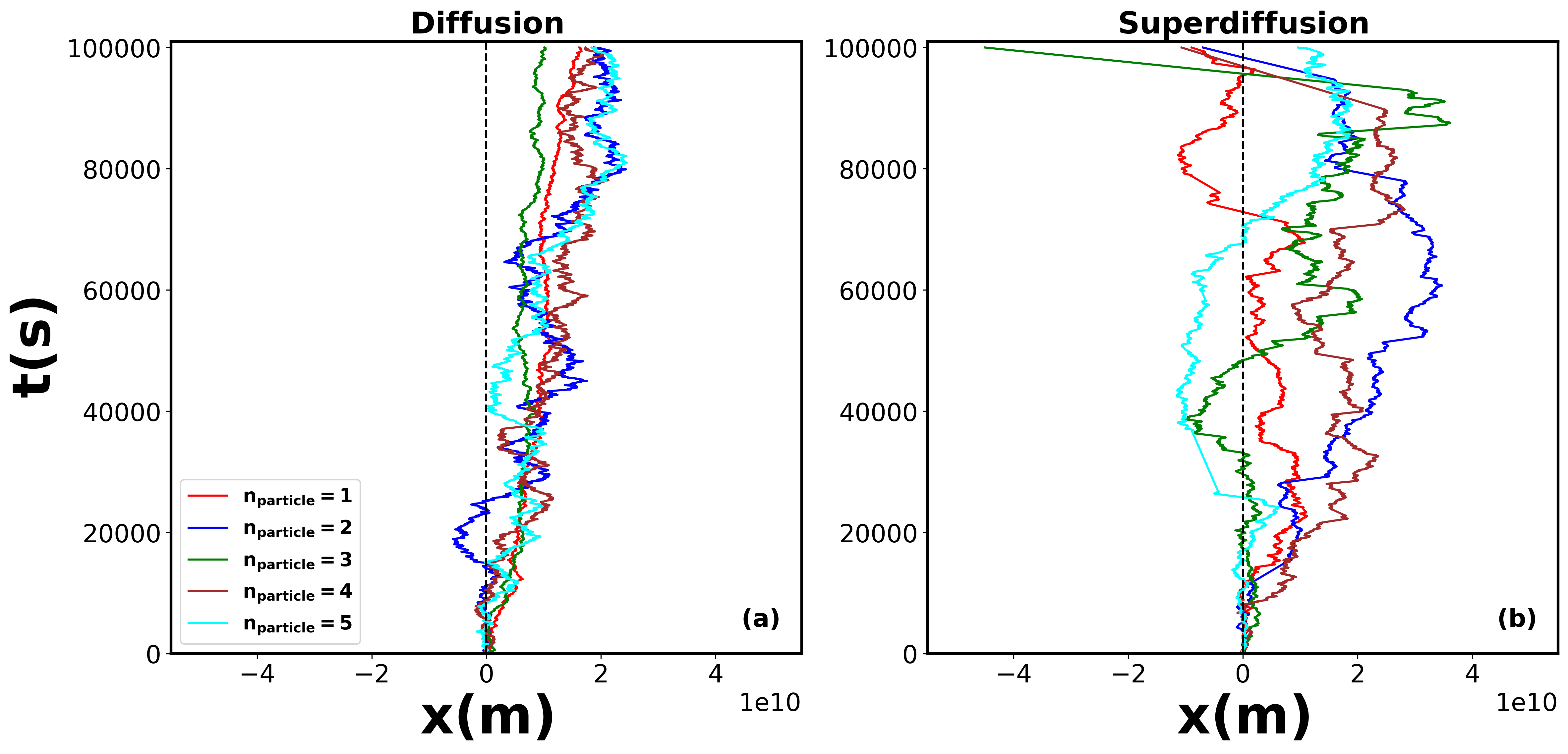}
\caption{Particle trajectories within a simulation with normal diffusion (panel (a)) and with superdiffusion (panel (b)).}
\label{traj-DSAvsSSA}
\end{figure}

Figure \ref{traj-DSAvsSSA} shows some typical particle's trajectories for the two transport cases. Several differences emerge from their comparison. Figure \ref{traj-DSAvsSSA}(a) displays five trajectories in the diffusive case. At early times particles remain confined near the shock. As time increases (up to 3000~s), advection becomes dominant, and the particles move downstream of the shock, with a lower probability of returning to the shock and gaining energy. Almost all the particles are fully advected downstream, except for a few.

Figure \ref{traj-DSAvsSSA}(b) shows the same trajectories for the superdiffusive case. At early times particles appear more confined near the shock region compared to the previous case. At later times, particles are able to cross the shock and return upstream. This behavior is due to the long free path displacements allowed by the L\'evy statistics, which increase the probability of returning to the upstream region. In Figure \ref{traj-DSAvsSSA}(b), long free paths in each particle's trajectory can be clearly observed. Even when a particle is almost completely advected downstream, it is still able to return upstream thanks to a long jump. This mechanism indeed increases the probability for particles to gain energy while crossing the shock and to speed up the acceleration process.

\begin{figure}[ht!]
\includegraphics[width=9cm]{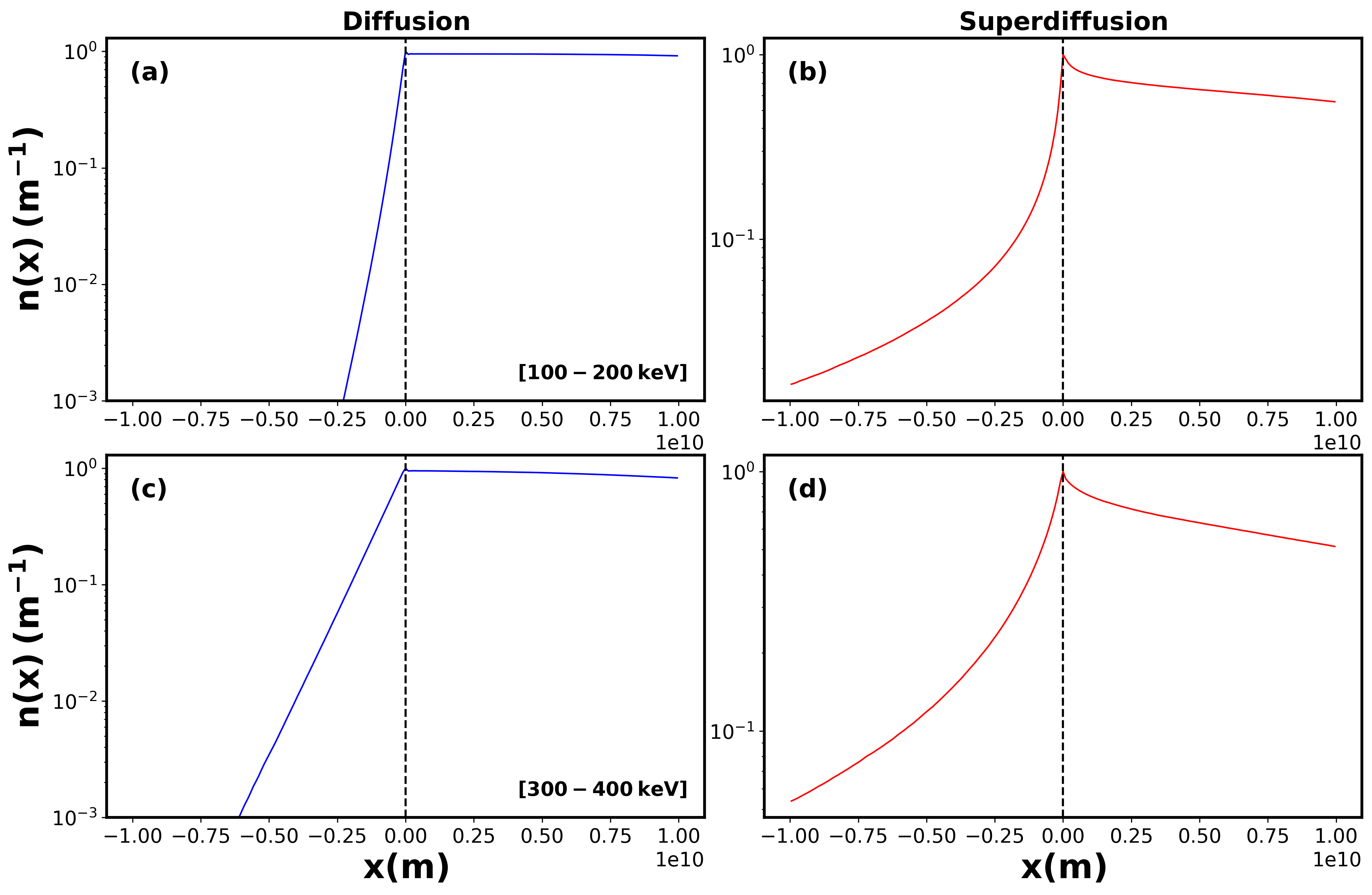}
\centering
\caption{
Panel(a): energetic particle density $n(x)$ for the 100-200 keV energy channel, in the case of normal diffusion with a scattering time $\tau_s$=50 s.  (Panel (b)): same quantity in the case of superdiffusion with $\mu=$ 2.5 and $t_0=$10s. (Panel (c)) and (Panel (d)): same format as top panels but for the 300-400 keV energy channel. 
}
\label{denscomp-100-400}
\end{figure}

Figure \ref{denscomp-100-400} displays the particle density computed as \citet{Prete2019, prete2021energetic} 
\begin{equation}
  n(x) = \Phi_0 \int_{0}^{\infty} P(x,t) dt, 
  \label{eq:integraleDens}
\end{equation}

where $\Phi_0$ represents the initial injected particle flux at the shock. Panels (a) and (c) show the density distributions for DSA for 100-200 keV and 300-400 keV particles, respectively. Quantities are normalized to the peak in the particle density profile. In both panels, the upstream region is characterized by an exponential decay, as expected. The slightly decreasing $n(x)$ downstream is related to a statistical problem with the number of particles escaping the simulation box in each energy bin (the higher the energy range the lower the number of particles in the box). The exponential decay in Figure \ref{denscomp-100-400} (c) is slower than in panel (a) because at higher energies, for a fixed scattering time, particles can travel a longer distances fixed by the diffusion length $L_{diff} = D_{||}/V_1$, where $D_{||}=\frac{1}{3} \lambda v$. Furthermore, there is a small peak in $n(x)$ in both channels related to the presence of particles moving around the shock. Figure \ref{denscomp-100-400} (b) and (d) show the same quantities, but in the case of superdiffusive transport. $n(x)$ does exhibit a similar trend in the two energy channels characterized by a power-law profile in the upstream region and a non-constant profile in the downstream region, as expected by the SSA theory (see \citet{Perri12a,Perrone13,Zimbardo13,Litvinenko14,Effenberger25}).

\subsection{Particle energy spectra}\label{sec:energyspectra}

To further study the acceleration process, we implement the calculation of the particle differential energy spectra, both for DSA and SSA, by counting particles in each energy bin. 
Indeed, particle energy spectra are expected to exhibit different slopes depending on whether diffusion or superdiffusion is considered \citep{kirk1996stochastic, Zimbardo13}. In particular, the differential energy spectrum, i.e.,  
\begin{equation}
\frac{dn}{dE}\propto E^{-\gamma},
\label{eq:spectraDown}
\end{equation}
has an exponent $\gamma$ that depends only on the shock compression ratio for DSA, while it depends also on the parameter $\mu$ in the case of superdiffusion. In particular, for non relativistic particles, we get \citep{Drury83,Zimbardo13}
{\large
\begin{equation}
\gamma = 
\begin{cases}
    \frac{2r+1}{2(r-1)} & \text{ DSA} \\
    \frac{3}{r-1} \frac{\mu-2}{\mu-1}  +1 & \text{ SSA}
\end{cases}
\label{eq:gammaDown}
\end{equation}
}

\begin{figure*}[!htbp]
\centering
\includegraphics[width=15cm]{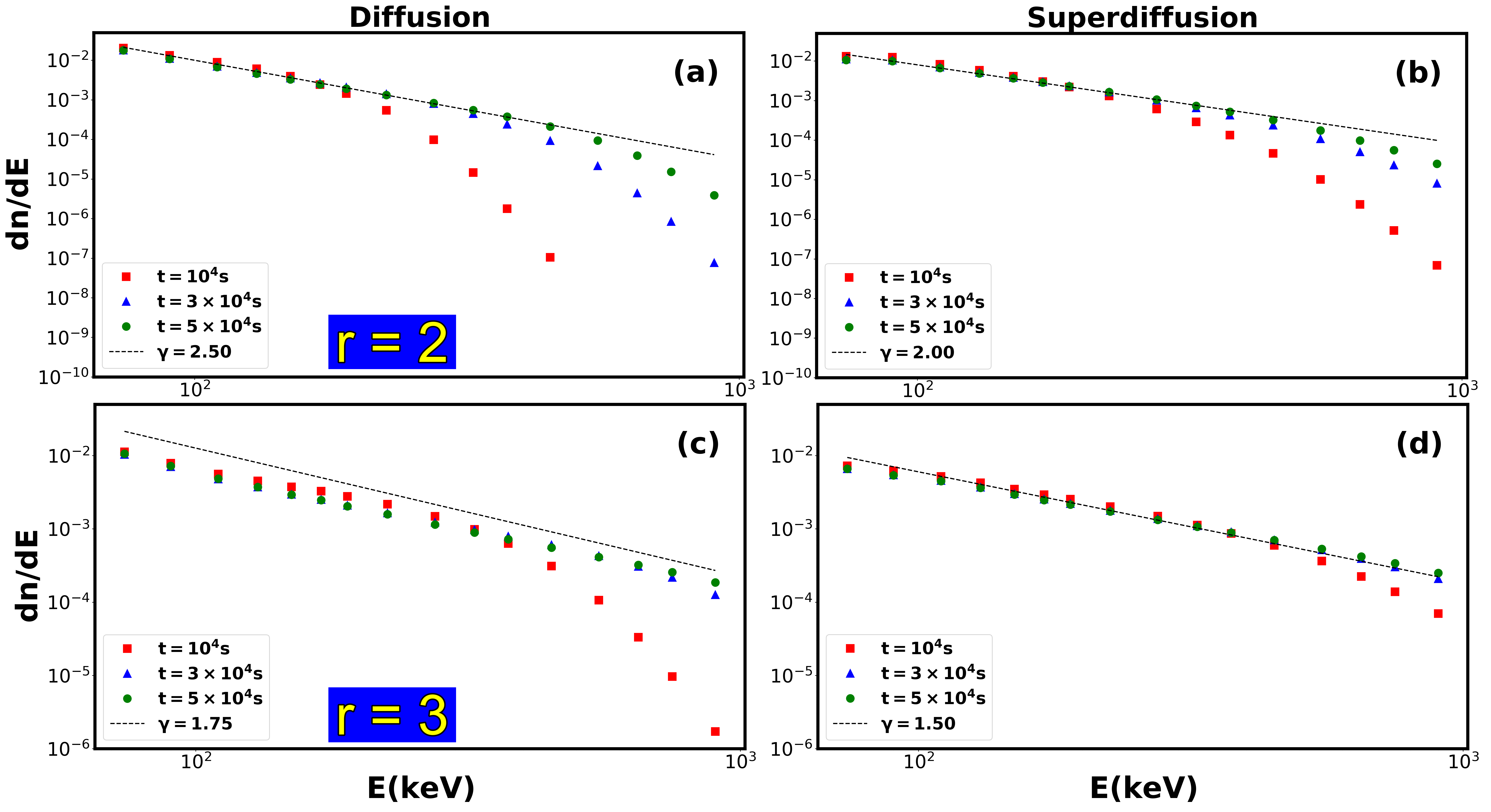}
\caption{
Energy spectra at different simulation times, for diffusion (panels (a) and (c)) and superdiffusion (panels (b) and (d)). In the top panels, the compression ratio is set to r = 2, while in the bottom panels r = 3. In the diffusive simulations the scattering time is $\tau_s$ =100 s (panel (a) and (c)). In the superdiffusive case (panels (b) and (d)) we set $\mu$=2.5 and $\tau_0=$ 10s.
}
\label{energySpectra}
\end{figure*}

\begin{figure*}[!htbp]
\centering
\includegraphics[width=15cm]{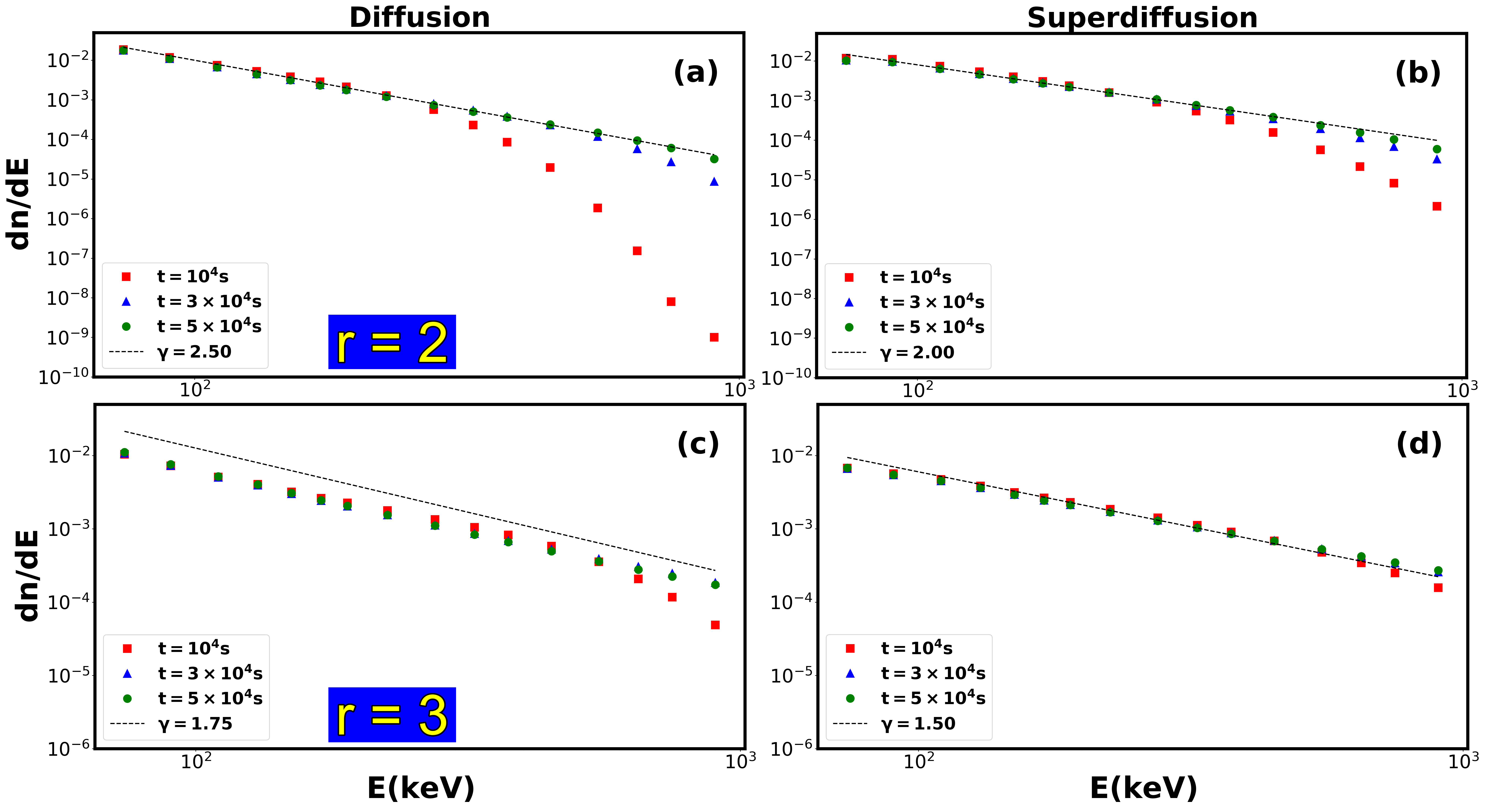}
\caption{
Same as in Figure~\ref{energySpectra}, but in this case we set the scattering time as $\tau_s=50$ s for the diffusive case, and we set the $\tau_0=5$ s for the superdiffusive case.
}
\label{fig:energySpectra2}
\end{figure*}

\begin{figure*}[!htbp]
\centering
\includegraphics[width=15cm]{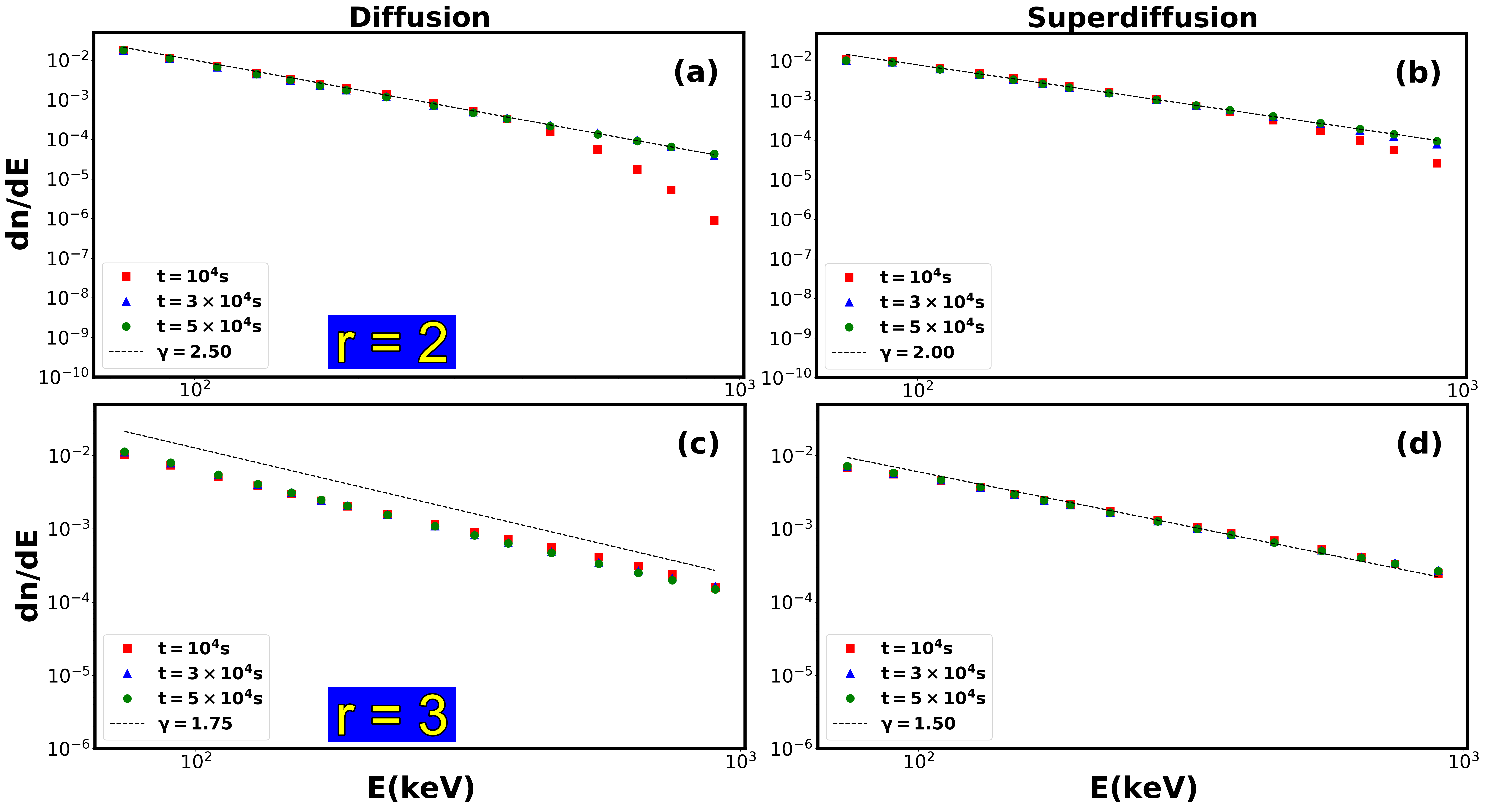}
\caption{
Same as in Figure~\ref{energySpectra}, but in this case we set the scattering time as $\tau_s=$ 25 s for the diffusive case, and $\tau_0=$ 2 s for the superdiffusive case.
}
\label{fig:energySpectra3}
\end{figure*}

A first check is to reproduce the above theoretical predictions within the test particle model. In order to do so, we run diffusive and superdiffusive simulations by varying the shock compression ratio. Figure~\ref{energySpectra} shows the energy spectra at different times (see figure legend) for $r=2$ (top panels) and $r=3$ (bottom panels). Also, the left panels refer to normal diffusion and the right panels to superdiffusion.  Particles are injected with an initial energy of $E_0$= 70~keV at $x=0$. The spectra have been computed by counting particles within 17 energy bins, ranging from 70~keV up to 1~MeV. We consider this energy range as appropriate for energetic particle measurements at many IP crossings. The figure legend also displays the theoretical values of the spectral index $\gamma$, as given by Eq. (\ref{eq:gammaDown}), for the chosen values of $r$ and $\mu$. 
As expected, as time goes on, more and more particles can be accelerated towards high energies, up to almost $1$ MeV both with diffusive and superdiffusive transport. At very late times, the differential energy spectra tend towards the predicted ones (dashed lines in the panels), thus confirming that our model can reproduce the theoretical expectations in both cases.\\
Another feature that emerges from Figure \ref{energySpectra} is that the energy spectra becomes broader and converge more rapidly to the theoretical prediction as the compression ratio increases, so that a larger compression ratio favors the acceleration process, as indicated by the momentum gain in Eq.~\ref{eq:momentum}; in addition, the presence of superdiffusive motion (panels (b) and (d)) seems to speed up the acceleration since the simulated curves reach the long time prediction earlier than the ones associated with a diffusive motion of particles.
The latter result is in agreement with \citet{Perri15}.

\begin{figure}[ht!]
\centering
\includegraphics[width=9cm]{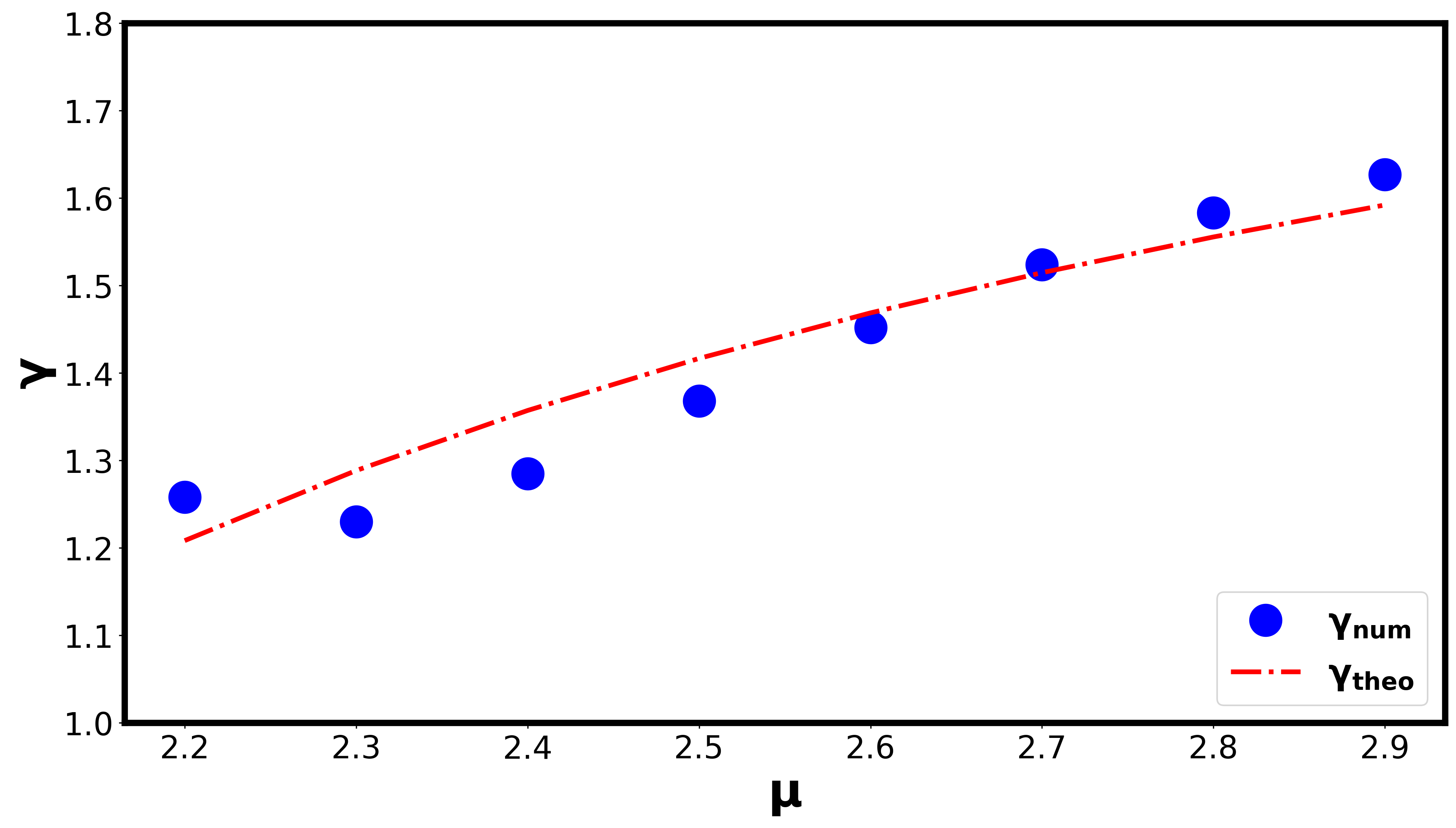}
\caption{Power-law exponent $\gamma$ in Eq.~\ref{eq:gammaDown} as a function of the superdiffusion parameter $\mu$. The blue dots correspond to values of $\gamma$ derived from the best power-law fits of the differential energy spectra from simulations with different $\mu$, whereas the red dash-dotted curve indicates its theoretical prediction (see Eq.~\ref{eq:gammaDown}). }
\label{fig:PlotGamma}
\end{figure}

Since the analysis of IP shock crossings indicates that the scattering time $\tau_s$ and the parameter $\tau_0$ may be smaller than those adopted in Figure~\ref{energySpectra}, we present the energy spectra for simulations with different values of $\tau_s$ and $\tau_0$ (see Figure~\ref{fig:energySpectra2} and \ref{fig:energySpectra3}). Following \citet{Giacalone12,prete2021energetic,Ding24}, we set $\tau_s = 50\,\mathrm{s}$ and $25 \,\mathrm{s}$ s for normal diffusion and $\tau_0 = 5\,\mathrm{s}$ and $2\,\mathrm{s}$ for superdiffusion. The spectral behaviour shown in Figure~\ref{fig:energySpectra2} and ~\ref{fig:energySpectra3} is similar to that in Figure~\ref{energySpectra}, except that acceleration is now faster. 
As the scattering times $\tau_s$ and $\tau_0$ decrease, the energy spectrum grows to higher energy values, since particles have a greater probability of being scattered,  returning to the shock and undergoing further energization, in agreement with theoretical predictions.\\
Figure~\ref{fig:PlotGamma} displays the energy spectrum exponent $\gamma$ defined in Eq.~\ref{eq:gammaDown} as a function of the parameter $\mu$ for a set of simulations. The blue dots represent the values of $\gamma$ obtained from the best power-law fits of the energy spectra of the simulations with superdiffusion, while the red dash-dotted line shows the theoretical prediction given by Eq.~\ref{eq:gammaDown}. We run simulations with $\mu$ ranging from 2.2 to 2.9. We do not show the $\mu=2.1$ case since particles can undergo very long jumps and a large ensemble of particles leaves the simulation box at very early simulation times, making the statistics poor for high energy particles. The numerical results are in reasonable agreement with the theoretical prediction. Consequently, the energy spectra become less steep as $\mu$ decreases, namely the particle mean square displacement grows more than linearly in time (see eq.(\ref{eq:meanSquareDispl})). 
It is worth mentioning that \citet{Aerdker25} used a L\'evy flight model to study the role of superdiffusion on particle acceleration at shocks. L\'evy flights are different from L\'evy walks since they represent a Markovian process in which, for a certain time $\tau$, a random walker can travel an arbitrarily large distance. This is at variance with the probability distribution of the free path lengths in Eq.~\ref{eq:levyWalk}, where a coupling between the length of the jump and the travel time does exist. In \citet{Aerdker25}, the inclusion of an  anomalous diffusion term in a 1D transport equation with energy gains has led to the emergence of harder energy spectra for particles and a faster acceleration than for both normal diffusion and L\'evy walk models. \citet{Aerdker25} ascribed this result to a lower escape probability that favors a more efficient acceleration and also to the fact that L\'evy flights do not have any limitation for large jump values, as for L\'evy walks, so that for the same anomalous diffusion exponent $\alpha$, particle energy spectra becomes harder allowing a quicker acceleration process at shocks.\\

\subsection{The ACE 14/12/2006 shock crossing}
\begin{figure}[ht!]
\centering
\includegraphics[width=9cm]{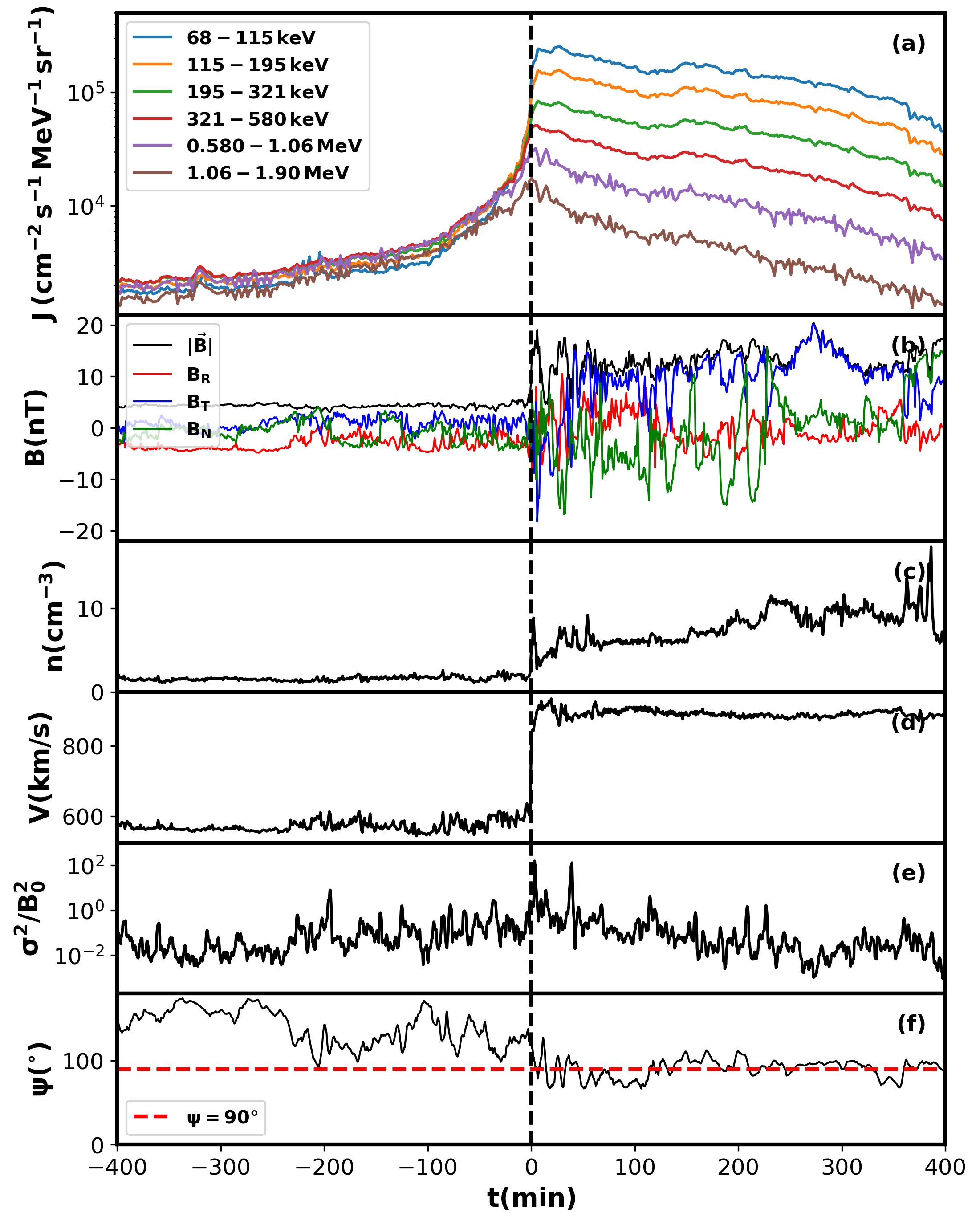}
\caption{Overview of the shock crossing observed by ACE on 14 December 2006. From top to bottom: (a) differential flux J of energetic particles measured by LEMS-120 in the energy channels from 68–115 keV to 1.06–1.9 MeV, see colors in the legend;  (b) magnetic field magnitude and its components at $1$ minute resolution; (c) proton number density; (d) bulk velocity in the spacecraft frame;  (e) magnetic variances normalized to the average magnetic field $B_0$, computed over the spatial scale corresponding to the gyroradius of a proton with 152 keV energy; 
(f) the angle $\psi$ between the magnetic field and the radial direction.}
\label{141206event}
\end{figure}

One of the objectives of this work is to directly compare the outcomes from the test particle model with spacecraft observations. We aim at reproducing the energetic particle fluxes in different energy channels upstream and downstream of the shock; having also added energy gains in the model, we can compare the differential energy spectra from the numerical experiment with the ones coming from satellite data. 
In particular, we consider the measurements of a shock crossing by ACE \citep{ACE1998} and we compute directly from the data the relevant parameters to be inserted into the numerical model. Figure~\ref{141206event} displays the 14 December 2006 shock event seen by the ACE spacecraft at 14:00 UT. The left hand side of the plot represents the upstream while the right hand side the downstream (notice the presence of the turbulent sheath region just behind the shock and of the flux rope at about 150 min downstream of the front). Panel (a) displays the energetic ion fluxes in six energy channels. Panel (b) shows the total magnetic field detected (in black) and its components in the Radial-Tangential-Normal (RTN) coordinate system, with a time cadence of ~1 minute. Panels (c) and (d) show the proton number density and the proton bulk speed, respectively. In panel (e), the total magnetic variance $\sigma^2=\sum_i \sigma_i^2$, computed via the variances of the magnetic field components $i=R,T,N$, namely $\sigma_i^2 = \langle (B_i - B_{i0})^2 \rangle_{\tau}$ (where $\langle \cdot \rangle_{\tau}$ represents a time average over a time scale $\tau$), is reported. $\sigma^2$ has been normalized to the square of the average magnetic field $B_0^2$ and computed separately for the upstream and downstream regions \citep{perri2012magnetic, zimbardo2020collisionless}. The time averages are calculated over time scales corresponding to the gyroradius of protons with energies corresponding to the midpoint energy channels listed in the legend of Figure~\ref{141206event} (a); in particular $\tau^{-1}=f_E=V_u/(2\pi \rho_E)$, where $V_u$ is the solar wind bulk speed upstream of the shock and $\rho_E$ is the Larmor radius of protons of energy $E$. Such variances are used as a proxy for the intensity of wave-particle scattering. In panel (e) the normalized variance computed at one time scale is reported. Finally, panel (f) displays the angle between the radial component of the magnetic field and the total magnetic field. We also determine the shock normal using the mixed method \citep{Paschmann2000, Trotta2023, Chiappetta2023}, applying it to intervals in the upstream and downstream regions of the shock, with averaging windows ranging from 1 minute to 5 minutes. We obtain the following shock parameters: the angle between the magnetic field and the shock normal $\theta_{B_n} = 44.6 \pm 22.7$; the gas compression ratio $r=3.4 \pm 0.9$; the plasma $\beta = 1.0 \pm 0.3$; the Alfv\'enic and magnetosonic Mach numbers of $M_A = 5.7 \pm 1.8$ and $M_{ms} = 4.2 \pm 1.3$, respectively. These parameters indicate that the shock is oblique and strong.

In the case of standard diffusive propagation, the flux determined at the observer position (i.e., $x=0$) is an exponential decay as a function of the time "distance" from the shock, namely
\begin{equation}
    J(0,E,t) \propto e^{-b|t-t_{sh}|},
    \label{eq:flux_diff}
\end{equation}
where the exponent $b$ is actually the inverse of the diffusion time $t_d$. As discussed in Section 3.1, the particle scattering time is directly related to the particle diffusion coefficient parallel to the field and since $D_{\parallel}$ can be directly computed from the diffusion coefficient along the radial direction, we can infer the scattering time directly from the analysis of the energetic particle fluxes upstream.
The diffusion coefficient component along the shock normal is given by $D_{xx}= t_d V_{1,sh} V_{sh}^{s/c}$, being $V_{1,sh}$ the upstream flow speed in the shock frame, while $V_{sh}^{s/c}$ is the shock speed in the spacecraft frame \citep{Giacalone12}.

When superdiffusion is taken into account, \citet{Perri07} have derived the particle flux at the satellite position far upstream of the shock, obtaining a
power-law decay in time for the particle flux
\begin{equation}
    J(0,E,t) \propto |t-t_{sh}|^{-\beta} \ ,
    \label{eq:flux_superdiff}
\end{equation}
where $\beta$ is related to the exponent of the mean square displacement as $\beta = 2 - \alpha $ (see Eq.~\ref{eq:meanSquareDispl}) \citep{Perri15}. As explained in Section \ref{sec:numerical}, $\alpha=4-\mu$, thus
the exponent of the scattering times in a L\'evy walk process (see equation (\ref{eq:scattSSA})) is directly linked to $\beta$, namely $\mu=\beta+2$. Since in the case of superdiffusion the scattering times depend on both $\mu$ and the parameter $\tau_0$, it is necessary to derive $\tau_0$ from observations in order to implement a data-driven run also for the superdiffusive model.

\begin{figure*}[ht!]
\sidecaption
\centering
\includegraphics[width=12cm]{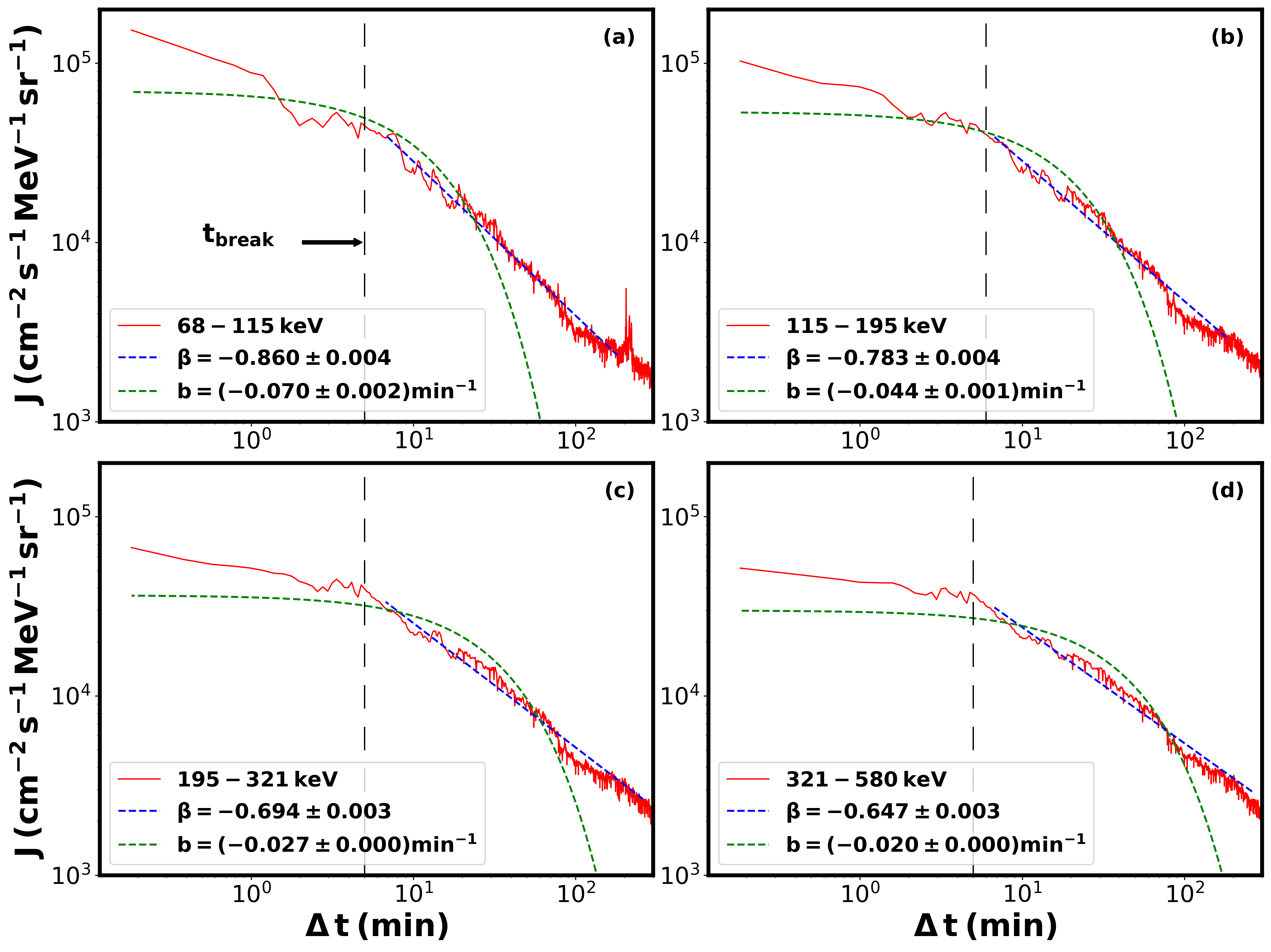}
\caption{Power-law (blue dashed line) and exponential fits (green dashed line) of the energetic particle fluxes (red line) in several energy channels (see legends) for the shock crossing of 14 December 2006. In the legends the exponent of the power-law fit, $\beta$, and of the exponential fit, $b$, are also indicated. The panels correspond to the energy channels (a) 68–115 keV, (b) 115-195 keV, (c) 195–321, and (d) 321–580 keV.}
\label{fits}
\end{figure*}

\begin{figure*}[ht!]
\sidecaption
\centering
\includegraphics[width=12.5cm]{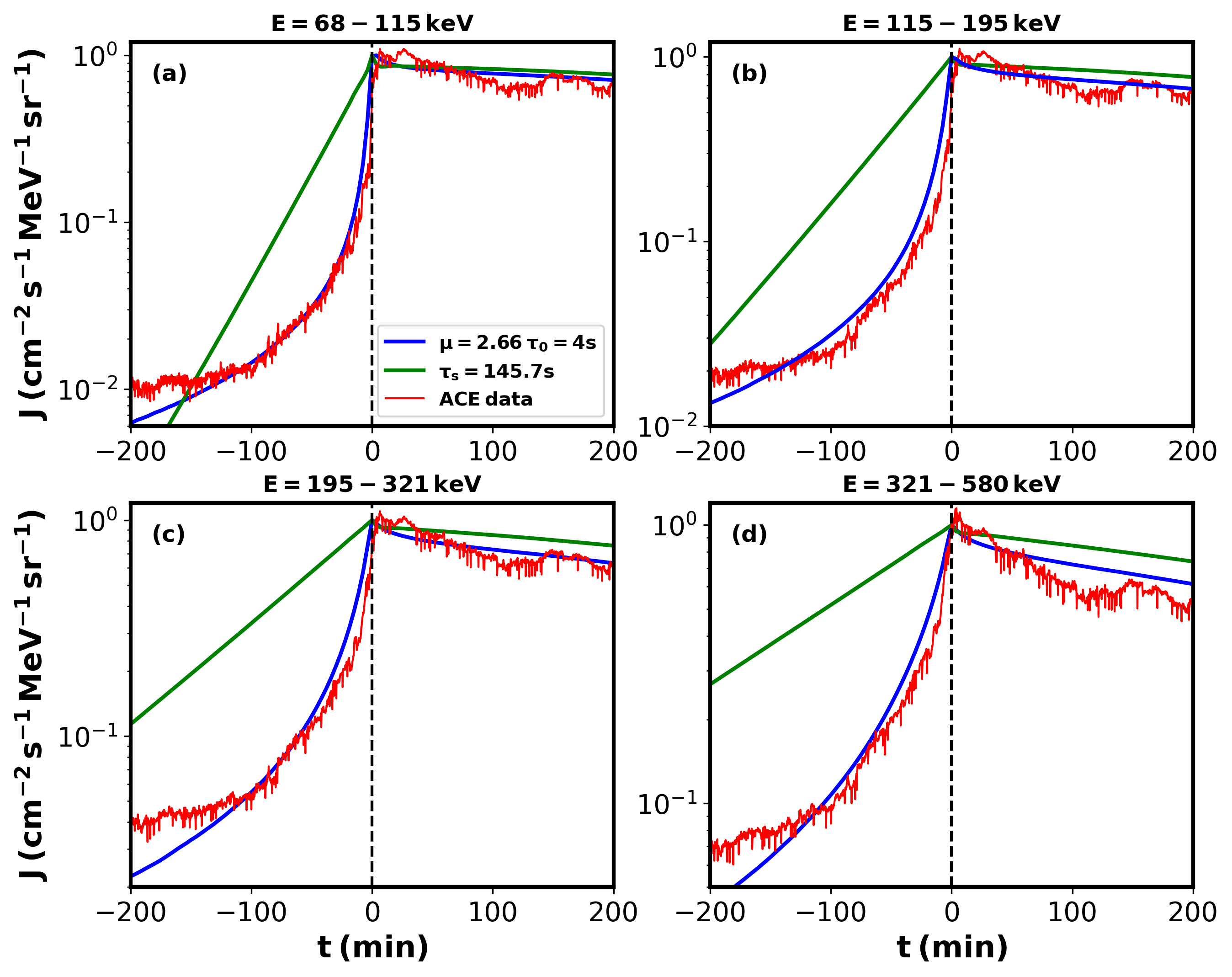}
\caption{Comparison between ACE measurements in several energy channels (red line) and the density profiles obtained from simulations in the case of diffusion (green line) and superdiffusion (blue line). }
\label{simulations}
\end{figure*}

According to \citet{Perri15b}, $\tau_0$ is related to the spacecraft crossing time $t_{break}$ at which the shape of the particle fluxes change from being a power law decay far upstream, to nearly flat closer to the shock front, namely
\begin{equation}
    \tau_0 = \biggl( \frac{v}{V_1^{sh}} \biggr)^\frac{\mu}{1-\mu} t_{break}.
    \label{eq:t0}  
\end{equation}
Notice that $t_{break}$ is determined by visual inspection of the particle flux profile, as reported below. 
Thus, in order to derive the model parameters from observations, we first analyze the particle fluxes in different energy channels upstream of the shock. We note that in our previous studies we found both normal diffusion and superdiffusion \citep{prete2021energetic}, depending on the energy channel and on the shock event.

It is clear from Figure~\ref{141206event}(a) that the upstream ion fluxes do not show the exponential decay predicted by DSA. Figure~\ref{fits} reports such upstream fluxes in log-log axis for four energy channels as a function of the time ``distance'' from the shock front: in each panel the red solid line represents the measured fluxes, while the green dashed line indicates the best exponential fit and the blue dashed line the power-law best-fit. The legend in each panel reports the values of the exponents found from the best-fits according to equations (\ref{eq:flux_diff}) and (\ref{eq:flux_superdiff}). A reduced $\tilde{\chi}^2$ has been computed for the diffusive and superdiffusive fits, and the results are shown in Table~\ref{tab:Table2}. 

\begin{table}[ht!]
\caption{Model comparison via reduced $\tilde{\chi}^2$.}
\label{tab:Table2}
\centering
\begin{tabular}{c c c }
\hline\hline
 Energy Channels (keV) & $\tilde{\chi}^2_{DSA}$ & $\tilde{\chi}^2_{SSA}$ \\
\hline
68-115 & 0.201 & 0.025  \\

115-195 & 0.152 & 0.026 \\

195-321 & 0.169 & 0.027 \\

321-580 & 0.167 & 0.029 \\
\hline
\end{tabular}
\tablefoot{
Reduced $\tilde{\chi}^2$ for all the energy channels computed for the best-fits of the fluxes using the diffusive and the superdiffusive models of the upstream fluxes.
}

\end{table}

The fitting procedure indicates that there is a good agreement between the data and the power law decay far upstream of the shock with an exponent $\beta <1$, indicating the presence of a superdiffusive transport for energetic particles. On the other hand, from the exponential fit of the particle fluxes, we get a parameter $b<0.1$ (see equation(\ref{eq:flux_diff})), resulting in a scattering time, according to Eq.~\ref{eq:scattering_times1}, between 80 and 200~s. 
We obtain a mean value of $\tau_s = 145.7$~s that has been used in the diffusive numerical simulations.

In each panel in Figure~\ref{fits} the vertical dashed line indicates the time at which the spacecraft detects a change in the energetic particle flux decay. Of course, higher time cadence measurements would lead to a better determination of such a time. From the best fit, we obtain 4 values for $\mu$ (directly from the determination of $\beta$) and 4 values for $\tau_0$. We take the mean of these values and we obtain $\mu$=2.66 and $\tau_0$= 4~s.

The values of $\tau_0$ are less than 10 s, in agreement with the events analyzed in \citet{Prete2019,zimbardo2020collisionless,prete2021energetic}. The two parameters $\mu$ and $\tau_0$ are given as input to the test particle superdiffusive code to implement data-driven simulations.

In Figure~\ref{simulations}, we present the comparison between the ACE fluxes in different energy channels and the output from numerical simulations, normalized to their maximum value. In each panel the red line represents the spacecraft detected fluxes, the green and the blue curves are the results of the simulations with diffusion and superdiffusion, respectively. The results clearly show that diffusive transport does not adequately reproduce the profiles observed by the spacecraft, whereas the superdiffusive case match the observed profiles very well, both in the upstream and in the downstream regions. On the other hand, we would like to remark that we use a very simplified model that neglects some physical aspects, such as possible variations of shock geometry, the possible amplification of turbulence close to the shock by streaming particles,  \citep{Ng03,Bell04,Perri23,Ha25}. In this connection, however, Figure~\ref{141206event} (e) shows that the level of turbulence estimated by $\sigma^2/B_0^2$, does not exhibit a  substantial increase upstream from -200 minutes towards the shock, suggesting that the transport properties do not change significantly with the shock distance and that there is no local amplification of the magnetic field fluctuations around the shock \citep{Perri12b}.

\begin{figure}[ht!]
\centering
\includegraphics[width=9cm]{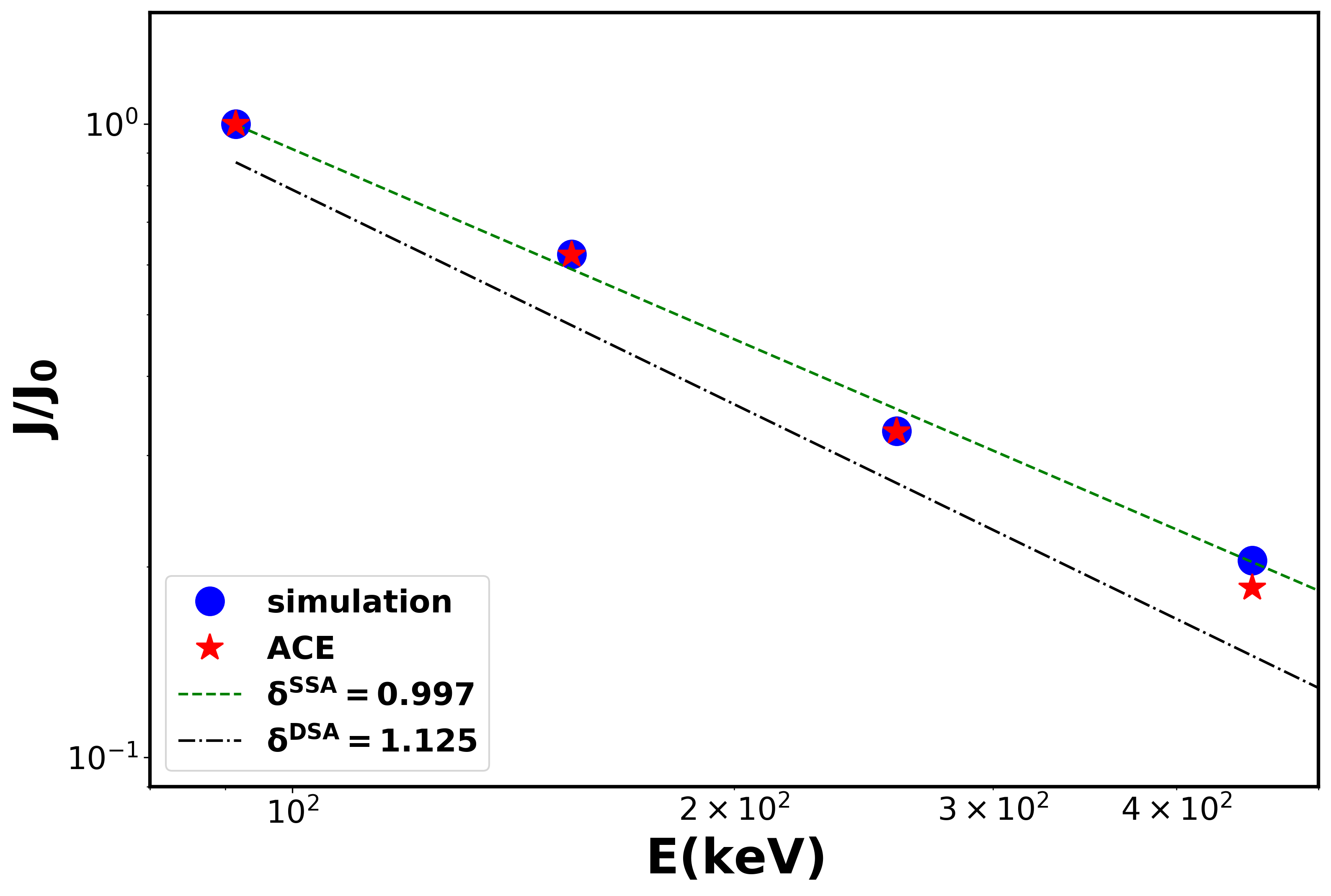}
\caption{Downstream energy spectra measured by ACE (red stars) and in numerical simulations with superdiffusion (blue dots). The values obtained for each energy channel are averaged over a 100-minutes time window. The green dashed lines represent the theoretical curves for $\delta$ in the case of superdiffusion. The black dotted line represents the theoretical value for the diffusion case. }
\label{Downstream_spectra}
\end{figure}

We further compute the differential flux, $dJ/dE \sim E^{-\delta}$, from ACE data and compare it with the one coming from test particle simulations with superdiffusion. Such a comparison is displayed in Figure~\ref{Downstream_spectra}. The values shown in the plot are averaged over a 100-minutes time window downstream (from the shock position to 100 minutes).  Such a choice of the time window has been dictated by the time series of the particle fluxes downstream that exhibit a slight increase for times greater than 100 min. The ratio $J/J_0$, where $J_0$ is a normalization factor obtained from the comparison between the simulations and the spacecraft data, is shown in the y-axis. The agreement between the red symbols (ACE data) and the blue symbols (simulations) is remarkable, thus leading to the conclusion that a very simplified model can capture observational features around IP shocks when superdiffusive transport is assumed to play the main role in the acceleration process. A slight discrepancy between the observed and the simulated spectra exists in the highest-energy channel, probably due to a statistical uncertainty in the high energy bins in both measurements and simulations.
Notice that the exponent of the differential flux, $\delta$, is directly related to the slope in Eq.~\ref{eq:gammaDown} of the differential energy spectrum, i.e.,  $\delta = \gamma - 1/2$. Figure~\ref{Downstream_spectra} further shows the expected values of the exponents $\delta$, in case of diffusion (black dotted-dashed line) and superdiffusion (green dashed line).
Both the observed and the simulated spectra lie within the values delimited by the theoretical predictions.

\section{Conclusions}
In this work we develop a new version of the codes presented in \citet{Prete2019,prete2021energetic} for the transport of particles around a collisionless shock. The difference with the previous version is related to the fact that particles now can be accelerated each time they cross the shock front. 
Such an acceleration process is studied under two different transport conditions for energetic particles: diffusion and superdiffusion. Since particles can reach different energies according to the number of times they cross the shock, an energy binning has been made at the end of the simulation for all the particles in the simulation box. This binning has been done by mimicking the Electron, Proton, and Alpha Monitor (EPAM)/LEMS-120 instrument on board ACE, so that a more straightforward comparison between observations and results from simulations can be made. 
We first compare the results obtained with diffusion and superdiffusion, setting the simulations parameters as follow: $\tau_s=100$ s, 50 s, and 25 s for the diffusion case, and $\mu = 2.5$, $\tau_0=10$ s, 5 s, and 2 s for the superdiffusive case. We select these parameters based on previous studies where similar values were derived from spacecraft observations \citep{Giacalone12, Kartavykh25, zimbardo2020collisionless,prete2021energetic}.
These choices are also supported by previous analyses of other shock events, which indicate that $\alpha$ ranges from 1.1 to 1.6 \citep{Perri15} for heliospheric shocks, implying $\mu = 4 - \alpha$ between 2.4 and 2.9. 
By comparing the diffusive and the superdiffusive simulations, we find that in the superdiffusive case, particles have a higher probability of crossing the shock multiple times and the effect of advection downstream at long times is reduced with respect to the diffusive case. In the latter case, only high energy particles can move easily upstream against advection, thus increasing their probability of crossing cross the shock front (and then to gain momentum).  

The acceleration process has been investigated by computing the differential energy spectra of particles at different running times. As the time goes on, the spectra measured in the simulations tend towards the theoretical prediction both for diffusive particles and for superdiffusive particles, with a more rapid convergence when the compression ratio of the shock is $r\sim 3$; this is due to the fact that at each shock crossing particles gain a fraction in momentum in the upstream/downstream frame that is proportional to the difference in plasma speeds.

Then, we study a shock crossing by the ACE spacecraft at 1~au that occurred on 14 December 2006. We analyze four ion energy channels and fit the upstream time profiles using both an exponential and a power-law decay. From the fitting procedure it is possible to derive the input parameters for the numerical simulations, namely the scattering time $\tau_s$ for DSA and the exponent $\mu$ and the value of $\tau_0$ of the power-law distribution of the scattering times for superdiffusion.
The best fit of the measured ion fluxes is obtained with a power-law decay with an exponent $\beta<1$, indicating superdiffusive transport for energetic ions. 
A comparison between numerical results and ACE observations indicate that such test particles model, in which particles perform a L\'evy random walk around the shock, can reproduce the observed energetic particle fluxes upstream and downstream of the shock. Superdiffusion has also been confirmed via the analysis of the particle energy spectrum during the ACE crossing in the downstream, which is fairly in agreement with the spectrum obtained by superdiffusive simulations with data-driven parameters. \\
This work sheds light on the possible role played by anomalous, superdiffusive transport, in the shock acceleration at IP and astrophysical shocks. Although we run a simple model that does not take into account variations with energies of the scattering times and then of the diffusion coefficients, we are able to reproduce some of the main observational features from in-situ data.
Indeed, in the present simulations particles always keep the same diffusive scattering time, even if they increase their energy while crossing the shock. A similar approximation is also made in the superdiffusive simulations: the parameter $\tau_0$ is actually dependent on the particle energy, but we fix it to a "typical" value in the simulation. This approximation  will be changed by fine tuning the scattering times as a function of energy in a future work.
Also, we would like to stress that this model does not include several physical effects such as the curvature of the shock surface, the influence of the magnetic field-shock normal angle, nor the full three-dimensional structure of the shock. In addition, being a test particle model, it neglects turbulence self-generated by particle-wave interactions near the shock, which may influence both particle transport and energization.
Further, the analysis of a large number of IP shocks is urgent in order to determine which shock conditions (namely, geometry, Mach numbers, turbulent ambient conditions upstream and downstream) can favor superdiffusive transport and the acceleration of high energy particles.
The recent Solar Orbiter shock database, described by \citet{Trotta25b} gives us the opportunity to carry out such an analysis.

\begin{acknowledgements}
GP, SP and GZ would like to thank Horst Fichtner, Sophie Aerdker,  Lukas Merten and Frederic Effenberger for their useful comments and suggestions. SP is grateful to J. le Roux, Du Toit Strauss, and J. Light for thoughtful discussions. SP and GZ acknowledge the project ‘Data-based predictions of solar energetic particle arrival to the Earth: ensuring space data and technology integrity from hazardous solar activity events’ (CUP H53D23011020001) ‘Finanziato dall’Unione europea – Next Generation EU’ PIANO NAZIONALE DI RIPRESA E RESILIENZA (PNRR) Missione 4 “Istruzione e Ricerca” - Componente C2 Investimento 1.1, ‘Fondo per il Programma Nazionale di Ricerca e Progetti di Rilevante Interesse Nazionale (PRIN)’ Settore PE09. Authors acknowledge the Space It Up project funded by the Italian Space Agency, ASI, and the Ministry of University and Research, MUR, under contract n. 2024-5-E.0 - CUP n. I53D24000060005. The authors further acknowledge support by the Italian PRIN 2022, project 2022294WNB, entitled “Heliospheric shocks and space weather: from multispacecraft observations to numerical modeling” (CUP H53D23000900006), Finanziato da Next Generation EU, fondo del Piano Nazionale di Ripresa e Resilienza (PNRR) Missione 4 “Istruzione e Ricerca” - Componente C2 Investimento 1.1, ‘Fondo per il Programma Nazionale di Ricerca e Progetti di Rilevante Interesse Nazionale (PRIN)’.
\end{acknowledgements}

\bibliographystyle{bibtex/aa}
\bibliography{bibtex/bibliography} 

\section*{Appendix A: L\'evy walk model}
Here we present the model used to reproduce the Le\'vy walks, which allows us to describe particle superdiffusive transport.
Starting from Equation~\ref{eq:levyWalk} we can determine the value of the normalization constant $C$, by fulfilling the condition
\begin{equation}
  \int_{0}^{+\infty} d\tau \int_{-\infty}^{+\infty} d\ell \, \Psi(\ell,\tau) = 1.
  \label{eq:cond_norm}
\end{equation}
\\
If we use the expression of $\Psi$ in Equation~\ref{eq:levyWalk}, also considering the simmetry of the probability distribution for positive and negative displacements $\ell$, we obtain
\[
C = \frac{1}{\tau_0} \biggl( \frac{\mu -1}{\mu} \biggr),
\]
where we make use of the relation $\tau_0=\ell_0/v$, and we set $2<\mu<3$ so that the integrals converge. 
\\
Let us write down the distribution function for the scattering times:
\[
\psi(\tau)  = \int_{-\infty}^{+\infty}  d\ell \, \Psi(\ell,\tau),
\]
which yields 
\\
\[
\psi (\tau) =
\left\lbrace
\begin{array}{ll}
C \;\;\;\;\;\;\;  \tau < \tau_0 \\
\\ 
C ({\tau}/{\tau_0})^{-\mu} \;\;\;\;\;\;\;  \tau > \tau_0 \, .
\end{array}
\right.
\]
We can generate $\psi(\tau)$ by extracting random numbers with probability distribution $f(\xi)$ and imposing the probability conservation, so that $f(\xi)d\xi=\psi(\tau)d\tau$. Since we deal with a uniform distribution of random number $\xi \in [0,1]$, we can set $f(\xi)d\xi \equiv C_\xi d\xi$. Normalizing $f(\xi)$ to one we find $C_\xi=1$, and we finally get 
\begin{equation}
    \xi = \int_{0}^{\tau_s} \psi(\tau) d\tau = 
    \left\lbrace
    \begin{array}{ll}
    C \,\tau_s \;\;\;\;\;\;\;  \tau_s < \tau_0 \\
    \\
    C\,\tau_0 + C \int_{\tau_0}^{\tau_s} (\frac{ \tau}{\tau_0})^{-\mu} \,d\tau, \;\;\;\;\;\;\;  \tau_s > \tau_0 \, .
    \end{array}
    \right.
    \label{eq:scattering_tauS}
\end{equation}
Solving the integral in Equation~\ref{eq:scattering_tauS}, we obtain:
\[
    \xi = \int_{o}^{\tau_s} \psi(\tau) d\tau = 
    \left\lbrace
    \begin{array}{ll}
    C \,\tau_s , \;\;\;\;\;\;\;  \tau < \tau_0 \\
    \\
    C\,\tau_0 + \frac{C}{\mu-1} \biggl(\frac{1}{\tau_0}\biggr)^{-\mu} \, \biggl[ \tau_0^{1-\mu}-\tau_s^{1-\mu} \biggr], \;\;\;\;\;\;\;  \tau > \tau_0 \, 
    \end{array}
    \right.
\]
where $\xi_0=C\,\tau_0=\frac{\mu-1}{\mu}$ separates the domain where $\psi(\tau)$ is constant from that where it is a power law. Inverting the above equation, we obtain the expression for the scattering time in the L\`evy walk framework for  $\xi>\xi_0$, as in Equation~\ref{eq:scattSSA}:
\[
 \tau_s = \tau_0 \left[\frac{1}{ \mu \left(1 - \xi \right)}\right]^\frac{1}{\mu - 1}.
\]
Clearly, $\tau_s = \xi/C$ for $\xi<\xi_0$. These relations allow us to determine the time required to perform a jump starting from a uniform distribution of $\xi$ ranging between 0 and 1. It is observed that, for $\xi \rightarrow 1$ long scattering times are obtained. Due to the specific space-time coupling in L\'evy walks, late times are required to observe long jumps.

\end{document}